# Inheritance of the exciton geometric structure from Bloch electrons in two-dimensional layered semiconductors

Jianju Tang[1], Songlei Wang[1], Hongyi Yu[1,2*]

[1] Guangdong Provincial Key Laboratory of Quantum Metrology and Sensing & School of Physics and Astronomy, Sun Yat-Sen University (Zhuhai Campus), Zhuhai 519082, China

[2] State Key Laboratory of Optoelectronic Materials and Technologies, Sun Yat-Sen University (Guangzhou Campus), Guangzhou 510275, China

* E-mail: yuhy33@mail.sysu.edu.cn

**Abstract:** We theoretically studied the exciton geometric structure in layered semiconducting transition metal dichalcogenides. Based on a three-orbital tight-binding model for Bloch electrons which incorporates their geometric structures, an effective exciton Hamiltonian is constructed and solved perturbatively to reveal the relation between the exciton and its electron/hole constituent. We show that the electron-hole Coulomb interaction gives rise to a non-trivial inheritance of the exciton geometric structure from Bloch electrons, which manifests as a valley-dependent center-of-mass anomalous Hall velocity of the exciton when two external fields are applied on the electron and hole constituents, respectively. The obtained center-of-mass anomalous velocity is found to exhibit a non-trivial dependence on the fields, as well as the wave function and valley index of the exciton. These findings can serve as a general guide for the field-control of the valley-dependent exciton transport, enabling the design of novel quantum optoelectronic and valleytronic devices.

## 1   Introduction

Atomically thin layers of semiconducting transition metal dichalcogenides (TMDs) have gained substantial interest owing to their potential as versatile platforms for exploring condensed matter phases and their promising applications in optoelectronic devices [1-5]. The direct band gap of monolayer TMDs is located at the hexagonal Brillouin zone corners, labelled as $\pm \mathbf{K}$ valleys. Due to the large effective masses at band edges and reduced dielectric screening in two-dimensional (2D) systems, these materials exhibit an exceptionally strong Coulomb interaction between charged carriers. As a result, the bound state of an electron-hole pair through the Coulomb interaction, called the exciton, plays a crucial role in photonic and optoelectronic properties of TMDs. The excitons can be viewed as a two-body system comprising a center-of-mass (CoM) motion and an electron-hole (e-h) relative motion. The e-h relative motion manifests as a discrete series of Rydberg states ($1s$, $2s$, $2p_\pm$, …) [6, 7], akin to the 2D hydrogen atom [8]. Excitons in different valleys of TMDs are endowed with valley optical selection rules [9-12], implying that the valley degree-of-freedom can be manipulated optically. In bilayer TMDs, excitons can be classified into intralayer and interlayer excitons [13], depending on whether the electron and hole reside in the same or different constituent monolayers. With the versatile tunability of monolayer and bilayer TMDs, the properties and dynamics of excitons can be tailored by various control knobs, such as in-plane electric fields, interlayer twisting, gate fields and so on [1-5].

In condensed matter systems, an intriguing aspect of the quasiparticle is the geometric structure which describes how its internal degree-of-freedom (spin, valley and orbital compositions, etc.) varies with parameters like position and momentum. It can give rise to geometric phases that have profound effects on various properties of the system [14]. In two-dimensional layered semiconductors, the variation of the orbital composition with momentum gives rise to non-trivial geometric structures for Bloch electrons in ±**K** valleys, which can be quantified by Berry curvatures in momentum space [15]. The geometric structure plays crucial roles in many exotic quantum phenomena including the valley orbital magnetic moments [16-18] and valley Hall effects [15, 19-24] where a Hall velocity transverse to the external electric field can emerge with a magnitude proportional to the Berry curvature. Such a Hall velocity doesn't require a magnetic field, thus is also called the anomalous velocity. As a composite quasiparticle, the exciton can inherit geometric structures from its Bloch electron and hole constituents which can manifest as an excitonic valley Hall effect as being reported in recent years [25-29]. Besides, previous studies have shown that geometric structures of the electron and hole constituents can lift the degeneracy of $2p_+$ and $2p_-$ exciton states and modify the energy spectrum [30-33] and optical properties [34-36]. In these previous works, the exciton geometric structure is often described by the direct summation of the Berry curvatures from the electron and hole constituents [30-33]. However, the strong Coulomb interaction can complicate the exciton's internal degree-of-freedom, and the bilayer stacking of TMDs brings further tunability to the electron and hole constituents. For a more rigorous analysis on the exciton geometric structure and its manifestation in quantum phenomena, a careful analysis on the internal structure involving the e-h relative motions needs to be carried out.

In this paper, we treat the electron and hole constituents of the exciton with the well-developed three-orbital model [37] where geometric structures of Bloch electrons are naturally incorporated. By utilizing Schrieffer-Wolff (SW) transformations and treating the inter-orbital coupling as a perturbation, we reveal an effective exciton Hamiltonian which contains not only the previously known energy correction terms, but also additional terms that couple different Rydberg states of the e-h relative motion. We found that these additional terms lead to a non-trivial inheritance of the exciton geometric structure from Bloch electrons. Such an exciton geometric structure can manifest as an anomalous CoM velocity under external fields, whose detailed form provides a general perspective for studying the exciton valley transport and designing novel valleytronic devices. In addition, the resultant anomalous CoM velocity of the exciton varies with both fields applied on the electron and hole constituents. We find that when external fields perturb the e-h relative motion (e.g., a homogeneous in-plane electric field), the resultant anomalous CoM velocity can be much larger than the case where only the CoM motion is perturbed (e.g., a thermal or density gradient field). Our work gives a rigorous derivation about how the exciton geometric structure in monolayer or bilayer TMDs is inherited from the Bloch electrons, which can serve as a guide for the field control of the valley-dependent exciton transport.

The paper is organized as follows. In section 2, we give an overview to the geometric structure of Bloch electrons and the CoM anomalous velocity of the exciton. In section 3, an effective Hamiltonian for excitons in TMDs is derived using a perturbative treatment, and the exciton geometric structure manifested as its CoM anomalous velocity under external fields is obtained. The last section is the summary and discussion.

## 2 The geometric structure of Bloch electrons

Near **K**, the three-orbital Hamiltonian of the Bloch electron involving $d_{x^2-y^2} - id_{xy}, d_{z^2}$

and $d_{x^2-y^2} + id_{xy}$ orbitals is [37]

$$\widehat{H}_e = \begin{pmatrix} \epsilon_r + \delta_r \widehat{\mathbf{p}}^2 & -\alpha \hat{p}_- & -\gamma \hat{p}_+ \\ -\alpha \hat{p}_+ & \epsilon_c + \delta_c \widehat{\mathbf{p}}^2 & \beta \hat{p}_- \\ -\gamma \hat{p}_- & \beta \hat{p}_+ & \epsilon_v + \delta_v \widehat{\mathbf{p}}^2 \end{pmatrix}, \qquad (1)$$

In the diagonal part, $\epsilon_r$, $\epsilon_c$ and $\epsilon_v$ give the **K**-point energies of the remote ($r$), conduction ($c$) and valence ($v$) bands, respectively. The off-diagonal terms represent the momentum-dependent inter-orbital couplings, with $\widehat{\mathbf{p}} \equiv -i\frac{\partial}{\partial \mathbf{r}_e} = -i\left(\frac{\partial}{\partial x_e}, \frac{\partial}{\partial y_e}\right)$ the momentum operator of the electron and $\hat{p}_\pm \equiv \hat{p}_x \pm i\hat{p}_y$. Without the external potential, $\widehat{\mathbf{p}}$ can be replaced by a classical number **p**, and we have kept up to the second-order (linear-order) of $\widehat{\mathbf{p}}$ in the diagonal (off-diagonal) terms. The resultant three bands of the Bloch electron are schematically shown in Fig. 1(a). The eigenstates $|u_{l,\mathbf{p}}\rangle$ of $\widehat{H}_e$ vary with **p** ($l = r, v, c$), which results in the internal geometric structure of the Bloch electron quantified by the Berry curvature $\Omega_{l,\mathbf{p}} \mathbf{e}_z = \frac{\partial}{\partial \mathbf{p}} \times i\langle u_{l,\mathbf{p}}|\frac{\partial u_{l,\mathbf{p}}}{\partial \mathbf{p}}\rangle$. Here $\mathbf{e}_{x/y/z}$ is the unit vector along $x/y/z$ direction. For a Bloch electron at **K**, its Berry curvature is $\Omega_r = 2\left(\frac{\gamma^2}{\epsilon_{rv}^2} - \frac{\alpha^2}{\epsilon_{rc}^2}\right)$, $\Omega_c = 2\left(\frac{\alpha^2}{\epsilon_{rc}^2} - \frac{\beta^2}{\epsilon_{cv}^2}\right)$ or $\Omega_v = 2\left(\frac{\beta^2}{\epsilon_{cv}^2} - \frac{\gamma^2}{\epsilon_{rv}^2}\right)$, with $\epsilon_{jl} \equiv \epsilon_j - \epsilon_l$ and $j, l = r, v, c$. Correspondingly, the hole Hamiltonian near $\tau'\mathbf{K}'$ is

$$\widehat{H}_h = \begin{pmatrix} -\epsilon_{r'} - \delta_{r'} \widehat{\mathbf{k}}^2 & \alpha' \hat{k}_{\tau'} & \gamma' \hat{k}_{-\tau'} \\ \alpha' \hat{k}_{-\tau'} & -\epsilon_{c'} - \delta_{c'} \widehat{\mathbf{k}}^2 & -\beta' \hat{k}_{\tau'} \\ \gamma' \hat{k}_{\tau'} & -\beta' \hat{k}_{-\tau'} & -\epsilon_{v'} - \delta_{v'} \widehat{\mathbf{k}}^2 \end{pmatrix}. \qquad (2)$$

Here $\widehat{\mathbf{k}} \equiv -i\frac{\partial}{\partial \mathbf{r}_h} = -i\left(\frac{\partial}{\partial x_h}, \frac{\partial}{\partial y_h}\right)$ is the momentum operator of the hole and $\hat{k}_\pm \equiv \hat{k}_x \pm i\hat{k}_y$. Note that for bilayer systems, the electron and hole can be in opposite layers with different parameters. $\tau' = +1$ ($\tau' = -1$) denotes the intravalley (intervalley) electron-hole pair. The hole Berry curvatures at $\tau'\mathbf{K}'$ are given by $\tau'\Omega_{r'} = 2\tau'\left(\frac{\alpha'^2}{\epsilon_{r'c'}^2} - \frac{\gamma'^2}{\epsilon_{r'v'}^2}\right)$, $\tau'\Omega_{c'} = 2\tau'\left(\frac{\beta'^2}{\epsilon_{c'v'}^2} - \frac{\alpha'^2}{\epsilon_{r'c'}^2}\right)$ and $\tau'\Omega_{v'} = 2\tau'\left(\frac{\gamma'^2}{\epsilon_{r'v'}^2} - \frac{\beta'^2}{\epsilon_{c'v'}^2}\right)$.

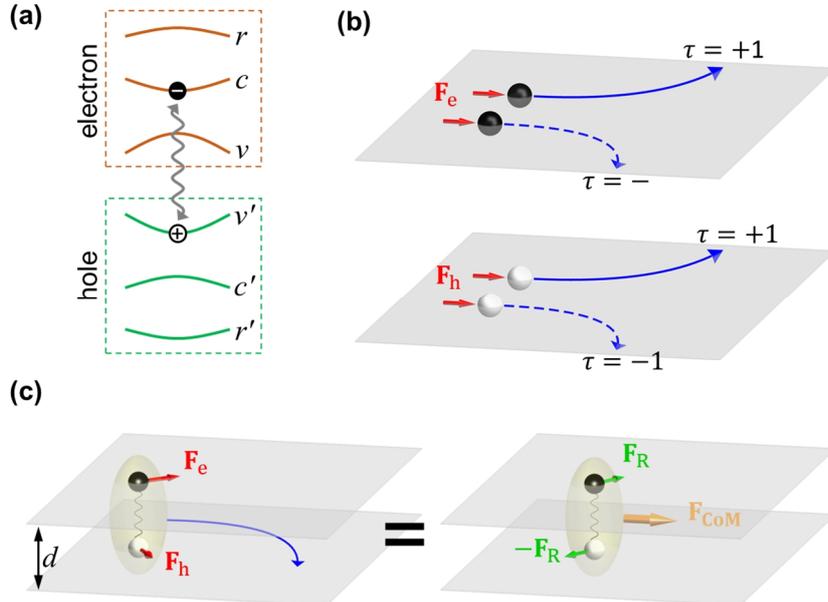

Figure 1. (a) The electron and hole bands near ±**K**. The wavy double arrow denotes the Coulomb interaction between them. (b) Schematic illustrations of valley Hall effects for the electron (upper panel) and hole (lower panel) under external forces $\mathbf{F}_e$ and $\mathbf{F}_h$, respectively. Blue solid (dashed) arrows denote the trajectories of carriers in $+\mathbf{K}$ ($-\mathbf{K}$) valley. (c) A schematic illustration of the Hall effect of the interlayer exciton when its electron and hole constituents experience external forces $\mathbf{F}_e$ and $\mathbf{F}_h$, respectively. The total force on the exciton can be decomposed into $\mathbf{F}_{\text{CoM}} \equiv \mathbf{F}_e + \mathbf{F}_h$ which perturbs the exciton CoM motion and $\mathbf{F}_R \equiv (m_h \mathbf{F}_e - m_e \mathbf{F}_h)/M$ which perturbs the e-h relative motion.

It is known that under an external potential $-\mathbf{F}_e \cdot \mathbf{r}_e$ ($-\mathbf{F}_h \cdot \mathbf{r}_h$) induced by a force field $\mathbf{F}_e$ ($\mathbf{F}_h$), the electron (hole) will acquire an anomalous transverse velocity $\Omega_l \mathbf{e}_z \times \mathbf{F}_e$ ($\tau' \Omega_{l'} \mathbf{e}_z \times \mathbf{F}_h$) which leads to the anomalous Hall effect [14] (Fig. 1(b)). An electron in the conduction band with an effective mass $m_e$ and a hole in the valence band with an effective mass $m_h$ can form an exciton through the Coulomb interaction, whose CoM motion experiences an external potential $-\mathbf{F}_{\text{CoM}} \cdot \mathbf{R}$ with $\mathbf{F}_{\text{CoM}} \equiv \mathbf{F}_e + \mathbf{F}_h$ the total force, $\mathbf{R} \equiv \frac{m_e \mathbf{r}_e + m_h \mathbf{r}_h}{M}$ the CoM coordinate and $M \equiv m_e + m_h$ the exciton mass. $\mathbf{F}_{\text{CoM}}$ then introduces a CoM anomalous velocity $\Omega_T \mathbf{e}_z \times \mathbf{F}_{\text{CoM}}$ to the exciton. The value $\Omega_T \equiv \frac{m_e^2}{M^2} \Omega_c + \frac{m_h^2}{M^2} \tau' \Omega_{v'}$, which is the sum of the electron and hole Berry curvatures (normalized by the effective mass related factors), can be viewed as the exciton Berry curvature under the effect of $\mathbf{F}_{\text{CoM}}$. However, the above rough treatment didn't fully take into account the effect of the Coulomb interaction $V(|\mathbf{r}_e - \mathbf{r}_h|)$, which adds an additional term $-\Omega_c \mathbf{e}_z \times \frac{\partial V(|\mathbf{r}_e - \mathbf{r}_h|)}{\partial \mathbf{r}_e}$ ($-\tau' \Omega_{v'} \mathbf{e}_z \times \frac{\partial V(|\mathbf{r}_e - \mathbf{r}_h|)}{\partial \mathbf{r}_h}$) to the electron (hole) anomalous velocity and can affect the exciton CoM anomalous velocity. Furthermore, for an interlayer exciton, $\mathbf{F}_e$ and $\mathbf{F}_h$ applied on the electron and hole constituents located in opposite layers can be tuned independently. Writing $\mathbf{F}_e \cdot \mathbf{r}_e + \mathbf{F}_h \cdot \mathbf{r}_h = \mathbf{F}_{\text{CoM}} \cdot \mathbf{R} + \mathbf{F}_R \cdot \mathbf{r}$ with $\mathbf{r} \equiv \mathbf{r}_e - \mathbf{r}_h$ the e-h relative coordinate and $\mathbf{F}_R \equiv \frac{m_h}{M} \mathbf{F}_e - \frac{m_e}{M} \mathbf{F}_h$ the force component that affects the e-h relative motion, the resultant CoM anomalous velocity should depend on both $\mathbf{F}_{\text{CoM}}$ and $\mathbf{F}_R$ thus cannot be solely determined by a single Berry curvature (see Fig. 1(c)).

In fact, writing the total potential as $U(\mathbf{r}_e, \mathbf{r}_h) \equiv V(|\mathbf{r}_e - \mathbf{r}_h|) - \mathbf{F}_e \cdot \mathbf{r}_e - \mathbf{F}_h \cdot \mathbf{r}_h$, the electron and hole velocity operators are in the forms:

$$\hat{\mathbf{v}}_e^{(c)} = \frac{\hat{\mathbf{p}}}{m_e} - \Omega_c \mathbf{e}_z \times \frac{\partial U(\mathbf{r}_e, \mathbf{r}_h)}{\partial \mathbf{r}_e},$$
$$\hat{\mathbf{v}}_h^{(v')} = \frac{\hat{\mathbf{k}}}{m_h} - \tau' \Omega_{v'} \mathbf{e}_z \times \frac{\partial U(\mathbf{r}_e, \mathbf{r}_h)}{\partial \mathbf{r}_h}. \quad (3)$$

The CoM velocity operator of the exciton is then $\hat{\mathbf{v}}_X = \frac{m_e}{M} \hat{\mathbf{v}}_e^{(c)} + \frac{m_h}{M} \hat{\mathbf{v}}_h^{(v')} = \hat{\mathbf{Q}}/M + \mathbf{v}_{eh} + \hat{\mathbf{v}}_{\text{int}}$. Here $\hat{\mathbf{Q}}/M$ is the trivial CoM group velocity, and $\mathbf{v}_{eh} + \hat{\mathbf{v}}_{\text{int}}$ is the exciton's CoM anomalous velocity operator induced by the electron and hole Berry curvatures, as can be seen from their equation forms

$$\mathbf{v}_{eh} \equiv \frac{m_e}{M} \Omega_c \mathbf{e}_z \times \mathbf{F}_e + \frac{m_h}{M} \tau' \Omega_{v'} \mathbf{e}_z \times \mathbf{F}_h = \mathbf{e}_z \times (\Omega_T \mathbf{F}_{\text{CoM}} + \delta\Omega \mathbf{F}_R),$$
$$\hat{\mathbf{v}}_{\text{int}} \equiv -\delta\Omega \mathbf{e}_z \times \frac{\partial V(r)}{\partial \mathbf{r}}. \quad (4)$$

Here $\delta\Omega \equiv \frac{m_e}{M}\Omega_c - \frac{m_h}{M}\tau'\Omega_{v'}$, and the values of $\Omega_{r/c/v}$, $\Omega_T$, $\delta\Omega$ are summarized in Table I. $\mathbf{v}_{eh}$ corresponds to the CoM sum of the electron and hole anomalous velocities, which is a constant independent on the exciton wave function. $\hat{\mathbf{v}}_{int}$ is the anomalous velocity operator introduced by the e-h Coulomb interaction. Setting the exciton basis as the Rydberg series $|nl\rangle$ of the traditional 2D hydrogen model and keeping only the four lowest-energy states $1s$, $2s$, $2p_\pm$, one can write

$$\hat{\mathbf{v}}_{int} \approx \sum_{n=1}^{2} (\langle 2p_+|\hat{\mathbf{v}}_{int}|ns\rangle|ns\rangle\langle 2p_+| + \langle 2p_-|\hat{\mathbf{v}}_{int}|ns\rangle|ns\rangle\langle 2p_-|) + \text{h.c.} \quad (5)$$

$\hat{\mathbf{v}}_{int}$ only has the off-diagonal terms since it changes the angular momentum quantum number by $\pm 1$. The CoM anomalous velocity $\mathbf{v}_a$ of a given exciton state equals the sum of $\mathbf{v}_{eh}$ and the expectation value $\langle \hat{\mathbf{v}}_{int}\rangle$. For $\langle \hat{\mathbf{v}}_{int}\rangle$ to be finite, the exciton must be in the coherent superposition of $|ns\rangle$ and $|2p_\pm\rangle$, which can be induced by $\mathbf{F}_R$ that affects the e-h relative motion or a possible coupling between the CoM and e-h relative motions.

Table 1. The electron Berry curvatures $\Omega_{r/c/v}$ at $\mathbf{K}$ calculated from the three-orbital model, and $\Omega_T \equiv (m_e^2\Omega_c + m_h^2\tau'\Omega_{v'})/M^2$, $\delta\Omega \equiv (m_e\Omega_c - m_h\tau'\Omega_{v'})/M$ for excitons in monolayer MoSe$_2$ (unit: nm$^2$). The hole Berry curvature is related to that of the electron through $\Omega_{r'/c'/v'} = -\Omega_{r/c/v}$ in monolayer TMDs.

| $\Omega_r$ | $\Omega_c$ | $\Omega_v$ | $\Omega_T$ ($\tau'=+1$) | $\Omega_T$ ($\tau'=-1$) | $\delta\Omega$ ($\tau'=+1$) | $\delta\Omega$ ($\tau'=-1$) |
|---|---|---|---|---|---|---|
| 0.023 | −0.125 | 0.148 | −0.072 | 0.030 | 0.036 | −0.138 |

## 3   The effective Hamiltonian and CoM anomalous velocities of excitons

Below we give a rigorous analysis to the exciton geometry structure and quantitatively calculate the corresponding CoM anomalous velocity under external forces. Starting from the three-orbital models of the electron and hole constituents (Eq. (1) and (2)), the full exciton Hamiltonian is

$$\hat{H} = \hat{H}_e \otimes \hat{I}_h + \hat{I}_e \otimes \hat{H}_h + U(\mathbf{r}_e, \mathbf{r}_h)\hat{I}_e \otimes \hat{I}_h, \quad (6)$$

where $\hat{I}_e$ and $\hat{I}_h$ are the 3x3 identity matrices in the electron and hole subspaces, respectively. Considering the presence of off-diagonal terms in $\hat{H}_e$ and $\hat{H}_h$, we apply two consecutive SW transformations $e^{\hat{S}'}e^{\hat{S}}\hat{H}e^{-\hat{S}}e^{-\hat{S}'}$ to perturbatively diagonalize $\hat{H}$. The first transformation involving the anti-hermitian operator $\hat{S}$ diagonalizes the non-interacting e-h pair $e^{\hat{S}}(\hat{H}_e \otimes \hat{I}_h + \hat{I}_e \otimes \hat{H}_h)e^{-\hat{S}}$ up to the second order of $\hat{\mathbf{p}}$ and $\hat{\mathbf{k}}$, but at the same time it also introduces finite but weak off-diagonal terms to $e^{\hat{S}}U(\mathbf{r}_e, \mathbf{r}_h)\hat{I}_e \otimes \hat{I}_h e^{-\hat{S}}$. The second transformation involving $\hat{S}'$ is to ensure that $e^{\hat{S}'}e^{\hat{S}}\hat{H}e^{-\hat{S}}e^{-\hat{S}'}$ becomes diagonal up to the first order of $U$ (see Appendix I for details about the two SW transformations). After the diagonalization, the exciton problem can be reduced to the subspace with the electron located in $c$-band and hole in $v'$-band. Meanwhile, the electron and hole velocity operators are $\hat{\mathbf{v}}_e \equiv e^{\hat{S}'}e^{\hat{S}}i[\hat{H}_e, \mathbf{r}_e]e^{-\hat{S}}e^{-\hat{S}'}$ and $\hat{\mathbf{v}}_h \equiv$

$e^{\hat{S}'}e^{\hat{S}}i[\hat{H}_h, \mathbf{r}_h]e^{-\hat{S}}e^{-\hat{S}'}$, respectively, whose components in $cv'$-subspace and the corresponding CoM anomalous velocity of the exciton are found to exactly coincide with Eq. (3) and (4), respectively.

By introducing the exciton CoM momentum $\hat{\mathbf{Q}} \equiv \hat{\mathbf{k}} + \hat{\mathbf{p}}$ and e-h relative momentum $\hat{\mathbf{q}} \equiv \frac{m_h}{M}\hat{\mathbf{p}} - \frac{m_e}{M}\hat{\mathbf{k}}$, we get the effective exciton Hamiltonian in $cv'$-subspace:

$$\hat{H}_{cv'} \approx \hat{H}_{X,0} + \delta\hat{H} + \hat{H}_C + \mathbf{e}_z \cdot \left(\frac{\Omega_T \mathbf{F}_{CoM} + \delta\Omega \mathbf{F}_R}{4} \times \hat{\mathbf{Q}}\right) - \mathbf{F}_{CoM} \cdot \mathbf{R}. \quad (7)$$

On the above right-hand-side, the effects of the last two terms are trivial and well-known: $\mathbf{e}_z \cdot \left(\frac{\Omega_T \mathbf{F}_{CoM} + \delta\Omega \mathbf{F}_R}{4} \times \hat{\mathbf{Q}}\right)$ introduces a $\mathbf{Q}$-dependent energy shift to the exciton, and $\mathbf{F}_{CoM} \cdot \mathbf{R}$ leads to the drift of the CoM momentum $\hat{\mathbf{Q}}$. Below we focus on the more essential first three terms $\hat{H}_{X,0}$, $\delta\hat{H}$ and $\hat{H}_C$.

The first term corresponds to the well-studied 2D hydrogen-like Hamiltonian

$$\hat{H}_{X,0} = \epsilon_{cv'} + \frac{\hat{\mathbf{Q}}^2}{2M} + \frac{\hat{\mathbf{q}}^2}{2\mu} + V(r), \quad (8)$$

with $\mu \equiv m_e m_h/M$ the reduced mass. The eigenstates of $\hat{H}_{X,0}$ are discrete Rydberg states $e^{i\mathbf{Q}\cdot\mathbf{R}}|nl\rangle$ with $nl = 1s, 2s, 2p_\pm$, whose CoM momentum $\mathbf{Q}$ and angular momentum $l = 0, 0, \pm1$ are good quantum numbers. Note that the nonlocal screening of the 2D layered geometry leads to a non-hydrogenic behavior for eigenstates of $\hat{H}_{X,0}$, and the resultant $2s$ exciton energy is higher than those of $2p_\pm$ [38-40], see Fig.2(a) inset.

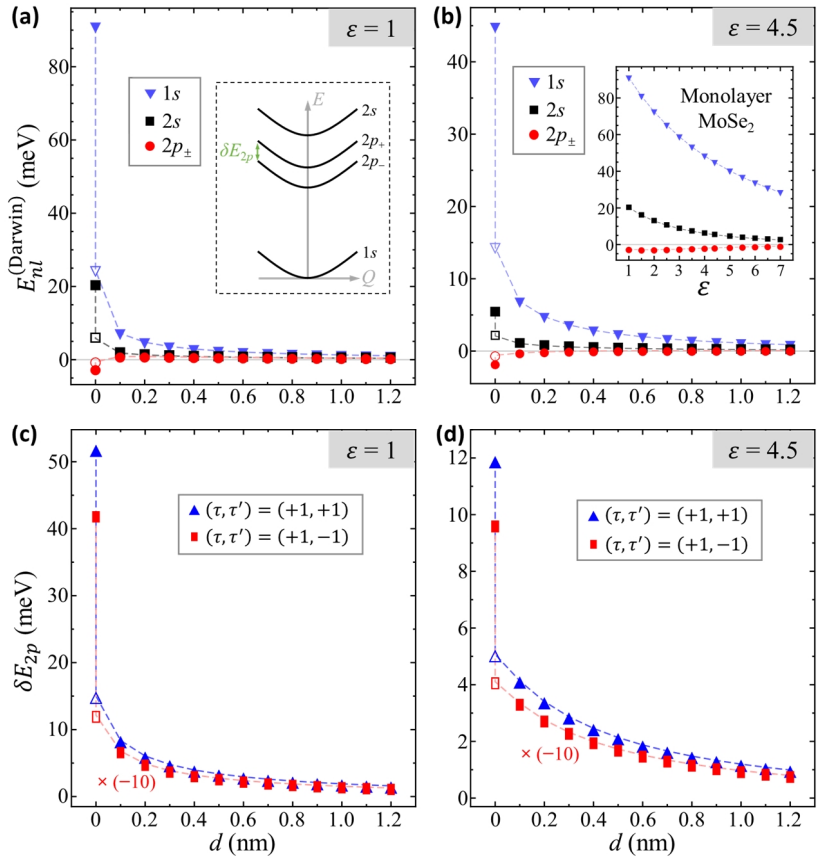

Figure 2. (a) The Darwin term induced energy correction $E_{nl}^{(Darwin)}$ to 1s, 2s, and $2p_{\pm}$ states of the interlayer exciton as a function of the interlayer distance $d$ in suspended MoSe$_2$ with environmental dielectric constant $\varepsilon = 1$. $d = 0$ corresponds to the monolayer case. The inset illustrates the energy alignment of 1s, 2s, $2p_+$ and $2p_-$ states. (b) $E_{nl}^{(Darwin)}$ for excitons in hBN-encapsulated MoSe$_2$ with $\varepsilon = 4.5$. The inset shows the dependence of $E_{nl}^{(Darwin)}$ on $\varepsilon$ in monolayer MoSe$_2$. (c) The SOC induced splitting $\delta E_{2p} \equiv E_{2p_+} - E_{2p_-}$ as a function of the interlayer distance $d$, for excitons in suspended MoSe$_2$. (d) $\delta E_{2p}$ for excitons in hBN-encapsulated MoSe$_2$. In (a-d), solid symbols at $d = 0$ correspond to excitons in monolayer MoSe$_2$, whereas empty symbols correspond to excitons in bilayer MoSe$_2$ with $d = 0$ (i.e., two monolayers vertically overlap).

The **Q**-independent term $\delta \hat{H}$ in Eq. (7) is given by
$$\delta \hat{H} = A \partial^2 V(r) + \frac{\Omega_c + \tau' \Omega_{v'}}{2} \left( \frac{\partial V(r)}{\partial \mathbf{r}} \times \hat{\mathbf{q}} \right) \cdot \mathbf{e}_z, \tag{9}$$

with $A = \frac{1}{2} \left( \frac{\alpha^2}{\epsilon_{rc}^2} + \frac{\beta^2}{\epsilon_{cv}^2} + \frac{\gamma'^2}{\epsilon_{r'v'}^2} + \frac{\beta'^2}{\epsilon_{c'v'}^2} \right)$. Here the first (second) term on the right-hand-side corresponds to the well-known Darwin (spin-orbit coupling) term in the 2D hydrogen model [8], which produces an energy correction $E_{nl}^{(Darwin)}$ ($E_{nl}^{(SOC)}$) to $|nl\rangle$ [31-33]. Different from the 3D hydrogen model where $E_{nl}^{(Darwin)}$ is finite only for s-type wave functions, in the 2D case it becomes finite for all Rydberg states with zero and nonzero angular momentums. Specifically, the energy shifts $E_{nl}^{(Darwin)}$ are independent on the valley index, and are the same for $2p_+$ and $2p_-$. On the other hand, $E_{nl}^{(SOC)}$ is finite only for states with nonzero angular momentums, with opposite values for $2p_+$ and $2p_-$. For a quantitative estimation, we have numerically solved the exciton wave functions $|nl\rangle$ for $nl = 1s, 2s, 2p_{\pm}$ in monolayer and bilayer MoSe$_2$, from which the energy corrections induced by Darwin and spin-orbit coupling (SOC) terms are obtained, see Appendix II for calculation details. Fig. 2(a) and 2(b) show $E_{nl}^{(Darwin)}$ in suspended (with environmental dielectric constant $\varepsilon = 1$) and hBN-encapsulated ($\varepsilon = 4.5$) MoSe$_2$, respectively. Fig. 2(c,d) are those for $\delta E_{2p} \equiv E_{2p_+}^{(SOC)} - E_{2p_-}^{(SOC)}$. For intravalley excitons with $\tau = +1$, our calculation gives $\delta E_{2p} \approx 12$ meV for intralayer exciton in monolayer MoSe$_2$ encapsulated by thick hBN, in agreement with the experimental observation [41]. For interlayer excitons in bilayer TMDs, we find that $\delta E_{2p}$ varies with the interlayer separation $d$ and is generally in the order of several meV. On the other hand, $\delta E_{2p}$ for intervalley excitons with $\tau = -1$ is negligibly small, see Fig. 2(c,d). The energy alignment of 1s, 2s, $2p_+$ and $2p_-$ after taking into account the above energy corrections is illustrated in Fig. 2(a) inset.

Unlike $\hat{H}_{X,0}$ and $\delta \hat{H}$ which conserve the angular momentum, $\hat{H}_C$ on the right-hand-side of Eq. (7) changes the angular momentum quantum number by $\pm 1$, with a form given by
$$\hat{H}_C = -\frac{\delta \Omega}{2} \left( \frac{\partial V(r)}{\partial \mathbf{r}} \times \hat{\mathbf{Q}} \right) \cdot \mathbf{e}_z + \mathbf{F}_R \cdot \mathbf{r} + \left( \frac{\Omega_c \mathbf{F}_e - \tau' \Omega_{v'} \mathbf{F}_h}{4} \times \hat{\mathbf{q}} \right) \cdot \mathbf{e}_z. \tag{10}$$

$\hat{H}_C$ then couples $|ns\rangle$ and $|2p_{\pm}\rangle$ states and introduces a finite $\langle \hat{v}_{int} \rangle$ component to the CoM anomalous velocity of the exciton (Eq. (4)). Below we focus on the CoM anomalous velocities of 1s and 2s states, as they can be detected through their radiative recombination. For small values of $\mathbf{Q}$ and $\mathbf{F}_{CoM/R}$, $\hat{H}_C$ leads to a perturbative coupling between $|ns\rangle$ and $|2p_{\pm}\rangle$ which gives rise to

$$\langle \hat{v}_{int} \rangle_{ns} = \left(\frac{1}{M+\Delta M_{ns}} - \frac{1}{M}\right)\mathbf{Q} - \delta\Omega \eta_{ns} \mathbf{e}_z \times \mathbf{F}_R \\ + \eta'_{ns} \mathbf{e}_z \times [\delta\Omega \mathbf{F}_{CoM} + (\Omega_c + \tau'\Omega_{v'})\mathbf{F}_R]. \quad (11)$$

On the right-hand-side of Eq. (11), the first term comes from the modification to the exciton dispersion from the $\hat{\mathbf{Q}}$-related term in $\hat{H}_C$ (Eq. (10)), thus can be viewed as a correction $\Delta M_{ns}$ to the exciton mass, with $\frac{1}{M+\Delta M_{ns}} - \frac{1}{M} = \frac{\delta\Omega^2}{4}\left[\frac{|\langle ns|\partial_- V|2p_+\rangle|^2}{E_{ns}-E_{2p+}} + \frac{|\langle ns|\partial_+ V|2p_-\rangle|^2}{E_{ns}-E_{2p-}}\right]$. The second term is an anomalous velocity originating from $\mathbf{F}_R \cdot \mathbf{r}$ in $\hat{H}_C$, with $\eta_{ns} \equiv \frac{\langle \psi_{2p+}|\hat{r}_+|\psi_{ns}\rangle\langle \psi_{ns}|\partial_- V|\psi_{2p+}\rangle}{E_{ns}-E_{2p+}} + \frac{\langle \psi_{2p-}|\hat{r}_-|\psi_{ns}\rangle\langle \psi_{ns}|\partial_+ V|\psi_{2p-}\rangle}{E_{ns}-E_{2p-}}$. The third term comes from SOC, with $\eta'_{ns} \equiv \frac{\delta\Omega}{8}\left[\frac{\langle \psi_{2p+}|\partial_+|\psi_{ns}\rangle\langle \psi_{ns}|\partial_- V|\psi_{2p+}\rangle}{E_{ns}-E_{2p+}} + \frac{\langle \psi_{2p-}|\partial_-|\psi_{ns}\rangle\langle \psi_{ns}|\partial_+ V|\psi_{2p-}\rangle}{E_{ns}-E_{2p-}}\right]$. $\eta_{ns}$ and $\eta'_{ns}$ are dimensionless parameters determined by the exciton wave function.

For an exciton with $\mathbf{Q} \approx \mathbf{0}$, we get the final form of its CoM anomalous velocity $\mathbf{v}_a = \mathbf{v}_{eh} + \langle \hat{v}_{int} \rangle_{ns}$, which can be written as

$$\mathbf{v}_a \approx \mathbf{e}_z \times [(\Omega_T + \eta'_{ns}\delta\Omega)\mathbf{F}_{CoM} + ((1-\eta_{ns})\delta\Omega + \eta'_{ns}(\Omega_c + \tau'\Omega_{v'}))\mathbf{F}_R]. \quad (12)$$

Eq. (12) represents the central result of this work. We can see that $\mathbf{v}_a$ depends on both $\mathbf{F}_{CoM}$ and $\mathbf{F}_R$ and varies with the exciton wave function through $\eta_{ns}$ and $\eta'_{ns}$. Fig. 3(a) shows our calculated values of $\eta_{ns}$ for excitons with different interlayer distances $d$, whose absolute values are ~ 2. The opposite signs of $\eta_{1s}$ and $\eta_{2s}$ come from the fact that $E_{1s} - E_{2p_{\pm}} < 0$ and $E_{2s} - E_{2p_{\pm}} > 0$. Both $\eta_{1s}$ and $\eta_{2s}$ have weak dependences on $d$ and $\varepsilon$, but are nearly independent on the valley index. Fig. 3(b) shows the mass correction $\Delta M_{1s}/M$ for 1s exciton, which reaches maximum ($\approx$ 5%) for the intralayer intervalley exciton under a weak environmental screening ($\varepsilon = 1$). The values of $\eta'_{ns}$ and $\Delta M_{2s}/M$ are found to be extremely small (~ 0.01 and ~ 0.01%, respectively) thus not shown.

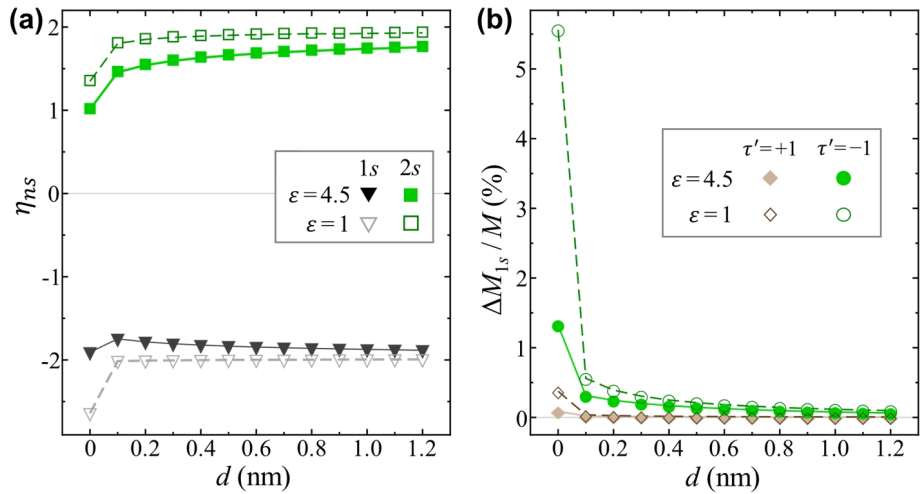

Figure 3. (a) The dimensionless parameter $\eta_{ns}$ for 1s and 2s exciton states as functions of the interlayer distance $d$, in the suspended ($\varepsilon = 1$) and hBN-encapsulated ($\varepsilon = 4.5$) MoSe$_2$. $\tau' = +1$ and $\tau' = -1$ have nearly the same $\eta_{ns}$ values. (b) The mass corrections $\Delta M_{1s}/M$ in the suspended and hBN-encapsulated TMDs as functions of $d$.

For interlayer excitons in bilayer TMDs, the electron and hole constituents located in opposite layers can experience different forces. This implies that the force fields $\mathbf{F}_{CoM}$ and $\mathbf{F}_R$ can be tuned independently. Below we consider three special combinations: (1) $\mathbf{F}_R = \mathbf{0}$ but $\mathbf{F}_{CoM} \neq \mathbf{0}$, which comes from the gradient force induced by an inhomogeneous density or thermal distribution, see the schematic illustration in Fig. 4(a); (2) $\mathbf{F}_{CoM} = \mathbf{0}$ but $\mathbf{F}_R \neq \mathbf{0}$, which corresponds to the case that a homogeneous in-plane electric field is applied on the exciton, see Fig. 4(b); (3) $\mathbf{F}_h = \mathbf{0}$ but $\mathbf{F}_e \neq \mathbf{0}$, that is, the force is applied only on the electron but not on the hole, see Fig. 4(c). The CoM anomalous velocity can be written as $\mathbf{v}_a = \Omega_{ns}^{(CoM/R/e)} \mathbf{e}_z \times \mathbf{F}_{CoM/R/e}$, with the three Berry curvatures given by

$$\Omega_{ns}^{(CoM)} = \Omega_T + \eta'_{ns}\delta\Omega,$$
$$\Omega_{ns}^{(R)} = (1 - \eta_{ns})\delta\Omega + \eta'_{ns}(\Omega_c + \tau'\Omega_{v'}),$$
$$\Omega_{ns}^{(e)} = \frac{m_e}{M}\Omega_c - \eta_{ns}\frac{m_h}{M}\delta\Omega + \eta'_{ns}\Omega_c. \qquad (13)$$

We have obtained $\Omega_{ns}^{(CoM/R/e)}$ for intralayer and interlayer excitons in MoSe$_2$, which are shown in Fig. 4(d-f). Because $\eta'_{ns} \approx 0$, $\Omega_{ns}^{(CoM)} \approx \Omega_T$ is nearly a constant independent on the exciton wave functions. However, $\Omega_{ns}^{(CoM)}$ exhibit very different values for intravalley and intervalley excitons, see Fig. 4(d). On the other hand, as shown in Fig. 4(e), the signs of $\Omega_{ns}^{(R)}$ are opposite for 1s and 2s interlayer excitons with the same valley index, implying that these two states exhibit opposite CoM anomalous velocities under the same in-plane electric field. Meanwhile the valley index can greatly affect the magnitude and sign of $\Omega_{ns}^{(R)}$. Interestingly, compared to $-\mathbf{F}_{CoM} \cdot \mathbf{R}$, the field component $-\mathbf{F}_R \cdot \mathbf{r}$ that affects the e-h relative motion has a more profound effect on the CoM anomalous velocity of the exciton, as can be seen from Fig. 4(d-f) that the maximum values of $\left|\Omega_{ns}^{(R)}\right|$ and $\left|\Omega_{ns}^{(e)}\right|$ are several times larger than that of $\left|\Omega_{ns}^{(CoM)}\right|$. We note that previous works [42, 43] have also pointed out that an in-plane electric field can generate a CoM anomalous velocity for the exciton, which corresponds to the second case in Eq. (13).

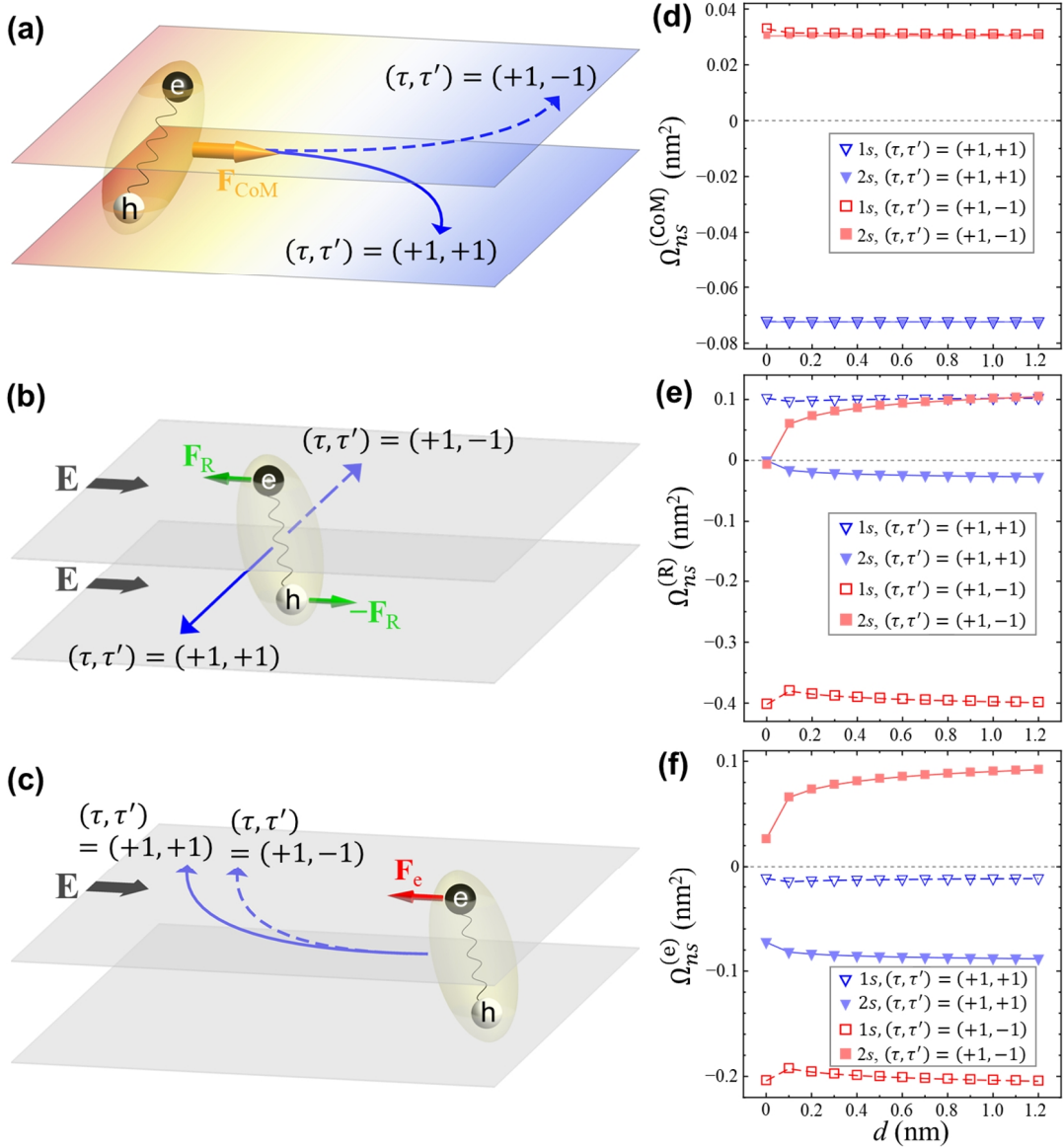

Figure 4. (a) A schematic illustration of the exciton valley Hall effect under $\mathbf{F}_R = 0$ but $\mathbf{F}_{CoM} \neq 0$ which is induced by a density or thermal gradient (color map). The orange arrow denotes the total force $\mathbf{F}_{CoM}$ applied on the exciton CoM motion. Solid and dashed blue arrows denote the trajectories of excitons with the valley indices $(\tau, \tau') = (+1, +1)$ and $(+1, -1)$, respectively, with their transverse motions induced by the CoM anomalous velocity of the exciton. (b) The exciton valley Hall effect under $\mathbf{F}_{CoM} = 0$ but $\mathbf{F}_R = -e\mathbf{E} \neq 0$ which is induced by a homogeneous in-plane electric field $\mathbf{E}$ (black arrows). Green arrows denote the electrostatic forces $\mathbf{F}_R$ and $-\mathbf{F}_R$ applied on the electron and hole constituents, respectively. (c) The exciton valley Hall effect under $\mathbf{F}_h = 0$ but $\mathbf{F}_e = -e\mathbf{E} \neq 0$ (the red arrow) which is from an electric field $\mathbf{E}$ (the black arrow) applied in the electron layer only. (d) The Berry curvature $\Omega_{ns}^{(CoM)}$ of $1s$ and $2s$ excitons as functions of the interlayer distance $d$ (see Eq. (13) in the maintext), which leads to the exciton transverse motion in (a). (e) The Berry curvature $\Omega_{ns}^{(R)}$ of $1s$ and $2s$ excitons, which leads to the exciton transverse motion in (b). (f) $\Omega_{ns}^{(e)}$ of $1s$ and $2s$ excitons, which leads to the exciton transverse motion in (c).

## 4 Summary and discussion

In conclusion, we have given a rigorous derivation of the exciton geometric structure in monolayer and bilayer TMDs, which is inherited from that of Bloch bands and can manifest as a CoM anomalous velocity of the exciton when external fields are applied on the electron and hole constituents. We have demonstrated that the CoM anomalous velocity has a non-trivial dependence on the two fields applied on the electron and hole, respectively, as well as the exciton wave function. A large CoM anomalous velocity can emerge even when the CoM motion of the exciton is not driven by the external fields. We have also calculated the energy corrections from the Darwin and SOC terms for intralayer and interlayer excitons in TMDs, which originate from geometric structures of Bloch bands. The obtained splitting between $2p_{\pm}$ exciton states agrees well with the experimental observation. We emphasize that our treatment is general and can be applied to excitons in other materials. For example, applying an interlayer bias can open a small band gap in bilayer graphene and induce large Berry curvatures for the carriers, which can result in a valley Hall effect as detected in experiments [44, 45]. Meanwhile, tunable excitons in bilayer graphene have also been observed experimentally [46]. This suggests that excitons in bilayer graphene can be another candidate to observe our proposed phenomena. The theoretical results in our work thus can serve as a guide for the field-control of the valley-dependent exciton transport, enabling the design of novel quantum optoelectronic and valleytronic devices.

In Eq. (12), the obtained CoM anomalous velocity is determined by the electron and hole Berry curvatures, suggesting that the exciton geometric structure is fully inherited from that of Bloch bands. Other geometric structures of excitons can also emerge, e.g., due to the position-dependent layer hybridizations in a bilayer moiré pattern [47] or CoM momentum dependent intervalley e-h exchange interaction [48]. In this work we do not consider the effect of moiré patterns. We also note that the strength of the e-h exchange interaction is proportional to the probability that the electron and hole spatially overlap, which is thus negligibly small for interlayer excitons due to the vertically separated electron and hole constituents. Meanwhile, the e-h exchange interaction can lead to the efficient valley depolarization (< 1 ps) for intravalley excitons in monolayer TMDs, which reverses the exciton's CoM anomalous velocity thus can diminish the valley Hall effect shown in Fig. 4. However, the e-h exchange interaction is negligible for intervalley excitons in monolayer TMDs and interlayer excitons in bilayer TMDs, resulting in their very long valley lifetimes ranging from ∼ 10 ns to ∼ μs [1, 13]. Also note that in Fig. 4 the largest magnitude of the CoM anomalous velocity corresponds to those of intervalley excitons with $\tau = -\tau'$. These properties make intervalley excitons very suitable for observing the proposed valley Hall effect. In monolayer TMDs, such intervalley excitons are momentum indirect thus cannot radiatively recombine directly. However, their radiative recombination can be assisted by impurity-scattering or phonon-emission, as experimentally observed in monolayer $WSe_2$ [49]. On the other hand, in bilayer TMDs with a 60° interlayer twist angle, intervalley excitons become momentum direct thus can radiatively recombine directly. The circularly polarized photon emissions of these intervalley excitons can facilitate the detection of the corresponding exciton valley Hall effect.

**Acknowledgement.** H.Y. acknowledges support by NSFC under grant No. 12274477, and

**Appendix I. Schrieffer-Wolff transformations on the exciton Hamiltonian**

We apply a Schrieffer-Wolff (SW) transformation $e^{\hat{S}}\hat{H}e^{-\hat{S}} = \hat{H} + [\hat{S},\hat{H}] + \frac{1}{2}[\hat{S},[\hat{S},\hat{H}]] + \cdots$ on the exciton Hamiltonian $\hat{H} = \hat{H}_e\otimes\hat{I}_h + \hat{I}_e\otimes\hat{H}_h + U(\mathbf{r}_e,\mathbf{r}_h)\hat{I}_e\otimes\hat{I}_h$, where the anti-hermitian operator $\hat{S}$ has a form

$$\hat{S} = \begin{pmatrix} 0 & -\frac{\alpha}{\epsilon_{rc}}\hat{p}_- & -\frac{\gamma}{\epsilon_{rv}}\hat{p}_+ \\ \frac{\alpha}{\epsilon_{rc}}\hat{p}_+ & 0 & \frac{\beta}{\epsilon_{cv}}\hat{p}_- \\ \frac{\gamma}{\epsilon_{rv}}\hat{p}_- & -\frac{\beta}{\epsilon_{cv}}\hat{p}_+ & 0 \end{pmatrix}\otimes\hat{I}_h + \hat{I}_e\otimes\begin{pmatrix} 0 & -\frac{\alpha'}{\epsilon_{r'c'}}\hat{k}_\tau & -\frac{\gamma'}{\epsilon_{r'v'}}\hat{k}_{-\tau} \\ \frac{\alpha'}{\epsilon_{r'c'}}\hat{k}_{-\tau} & 0 & \frac{\beta'}{\epsilon_{c'v'}}\hat{k}_\tau \\ \frac{\gamma'}{\epsilon_{r'v'}}\hat{k}_\tau & -\frac{\beta'}{\epsilon_{c'v'}}\hat{k}_{-\tau} & 0 \end{pmatrix}. \quad (A1)$$

The resultant non-interacting e-h pair Hamiltonian $\widetilde{H}_{eh} \equiv e^{\hat{S}}(\hat{H}_e\otimes\hat{I}_h + \hat{I}_e\otimes\hat{H}_h)e^{-\hat{S}}$ becomes diagonal and has the following form

$$\widetilde{H}_{eh} \approx \begin{pmatrix} \epsilon_r + \frac{\hat{\mathbf{p}}^2}{2m_r} & 0 & 0 \\ 0 & \epsilon_c + \frac{\hat{\mathbf{p}}^2}{2m_c} & 0 \\ 0 & 0 & \epsilon_v + \frac{\hat{\mathbf{p}}^2}{2m_v} \end{pmatrix}\otimes\hat{I}_h$$

$$+ \hat{I}_e\otimes\begin{pmatrix} -\epsilon_{r'} + \frac{\hat{\mathbf{k}}^2}{2m_{r'}} & 0 & 0 \\ 0 & -\epsilon_{c'} + \frac{\hat{\mathbf{k}}^2}{2m_{c'}} & 0 \\ 0 & 0 & -\epsilon_{v'} + \frac{\hat{\mathbf{k}}^2}{2m_{v'}} \end{pmatrix}. \quad (A2)$$

Here $m_r \equiv \frac{1}{2}\left(\delta_r + \frac{\alpha^2}{\epsilon_{rc}} + \frac{\gamma^2}{\epsilon_{rv}}\right)^{-1}$, $m_c \equiv \frac{1}{2}\left(\delta_c + \frac{\beta^2}{\epsilon_{cv}} - \frac{\alpha^2}{\epsilon_{rc}}\right)^{-1}$ and $m_v \equiv \frac{1}{2}\left(\delta_v - \frac{\beta^2}{\epsilon_{cv}} - \frac{\gamma^2}{\epsilon_{rv}}\right)^{-1}$ correspond to the effective masses of the three electron Bloch bands, and $m_{r'} \equiv -\frac{1}{2}\left(\delta_{r'} + \frac{\alpha'^2}{\epsilon_{r'c'}} + \frac{\gamma'^2}{\epsilon_{r'v'}}\right)^{-1}$, $m_{c'} \equiv -\frac{1}{2}\left(\delta_{c'} + \frac{\beta'^2}{\epsilon_{c'v'}} - \frac{\alpha'^2}{\epsilon_{r'c'}}\right)^{-1}$, $m_{v'} \equiv \frac{1}{2}\left(\frac{\beta'^2}{\epsilon_{c'v'}} + \frac{\gamma'^2}{\epsilon_{r'v'}} - \delta_{v'}\right)^{-1}$ are those of the hole bands. In Eq. (A2), we have kept up to the second-order of $\hat{\mathbf{p}}$ and $\hat{\mathbf{k}}$ for the diagonal terms, whereas the off-diagonal terms are at least second-order of $\hat{\mathbf{p}}$ and $\hat{\mathbf{k}}$ thus are all dropped because of the large inter-band energy separations.

The total potential is transformed to $\widetilde{U} \equiv e^{\hat{S}}U(\mathbf{r}_e,\mathbf{r}_h)\hat{I}_e\otimes\hat{I}_h e^{-\hat{S}}$ which has the below form:

$$\widetilde{U} \approx U(\mathbf{r}_e,\mathbf{r}_h)\hat{I}_e\otimes\hat{I}_h + \begin{pmatrix} \hat{H}_{\text{Darwin}}^{(r)} + \hat{H}_{\text{SOC}}^{(r)} & i\frac{\alpha}{\epsilon_{rc}}\partial_{e,-}U & i\frac{\gamma}{\epsilon_{rv}}\partial_{e,+}U \\ -i\frac{\alpha}{\epsilon_{rc}}\partial_{e,+}U & \hat{H}_{\text{Darwin}}^{(c)} + \hat{H}_{\text{SOC}}^{(c)} & -i\frac{\beta}{\epsilon_{cv}}\partial_{e,-}U \\ -i\frac{\gamma}{\epsilon_{rv}}\partial_{e,-}U & i\frac{\beta}{\epsilon_{cv}}\partial_{e,+}U & \hat{H}_{\text{Darwin}}^{(v)} + \hat{H}_{\text{SOC}}^{(v)} \end{pmatrix}\otimes\hat{I}_h$$

$$+ \hat{I}_e\otimes\begin{pmatrix} \hat{H}_{\text{Darwin}}^{(r')} + \hat{H}_{\text{SOC}}^{(r')} & i\frac{\alpha'}{\epsilon_{r'c'}}\partial_{h,\tau}U & i\frac{\gamma'}{\epsilon_{r'v'}}\partial_{h,-\tau}U \\ -i\frac{\alpha'}{\epsilon_{r'c'}}\partial_{h,-\tau}U & \hat{H}_{\text{Darwin}}^{(c')} + \hat{H}_{\text{SOC}}^{(c')} & -i\frac{\beta'}{\epsilon_{c'v'}}\partial_{h,\tau}U \\ -i\frac{\gamma'}{\epsilon_{r'v'}}\partial_{h,\tau}U & i\frac{\beta'}{\epsilon_{c'v'}}\partial_{h,-\tau}U & \hat{H}_{\text{Darwin}}^{(v')} + \hat{H}_{\text{SOC}}^{(v')} \end{pmatrix}. \quad (A3)$$

Here $\partial_{e/h,\pm} U \equiv \frac{\partial U}{\partial x_{e/h}} \pm i\frac{\partial U}{\partial y_{e/h}}$ and $\partial^2_{e/h} U \equiv \frac{\partial^2 U}{\partial x^2_{e/h}} + \frac{\partial^2 U}{\partial y^2_{e/h}}$. $\widehat{H}^{(r)}_{\text{Darwin}} \equiv \frac{1}{2}\left(\frac{\alpha^2}{\epsilon^2_{rc}} + \frac{\gamma^2}{\epsilon^2_{rv}}\right)\partial^2_e U$, $\widehat{H}^{(c)}_{\text{Darwin}} \equiv \frac{1}{2}\left(\frac{\alpha^2}{\epsilon^2_{rc}} + \frac{\beta^2}{\epsilon^2_{cv}}\right)\partial^2_e U$ and $\widehat{H}^{(v)}_{\text{Darwin}} \equiv \frac{1}{2}\left(\frac{\gamma^2}{\epsilon^2_{rv}} + \frac{\beta^2}{\epsilon^2_{cv}}\right)\partial^2_e U$ correspond to the well-known Darwin terms for electrons in $r$-, $c$- and $v$-bands, respectively. $\widehat{H}^{(r/c/v)}_{\text{SOC}} \equiv \frac{i}{4}\Omega_{r/c/v}\left(\partial_{e,-} U\hat{p}_+ - \partial_{e,+} U\hat{p}_-\right)$ are the spin-orbit coupling terms of the three electron bands. $\widehat{H}^{(r')}_{\text{Darwin}} \equiv \frac{1}{2}\left(\frac{\alpha'^2}{\epsilon^2_{r'c'}} + \frac{\gamma'^2}{\epsilon^2_{r'v'}}\right)\partial^2_h U$, $\widehat{H}^{(c')}_{\text{Darwin}} \equiv \frac{1}{2}\left(\frac{\alpha'^2}{\epsilon^2_{r'c'}} + \frac{\beta'^2}{\epsilon^2_{c'v'}}\right)\partial^2_h U$, $\widehat{H}^{(v')}_{\text{Darwin}} \equiv \frac{1}{2}\left(\frac{\gamma'^2}{\epsilon^2_{r'v'}} + \frac{\beta'^2}{\epsilon^2_{c'v'}}\right)\partial^2_h U$ and $\widehat{H}^{(r'/c'/v')}_{\text{SOC}} \equiv \frac{i}{4}\tau\Omega_{r'/c'/v'}\left(\partial_{h,-} U\hat{k}_+ - \partial_{h,+} U\hat{k}_-\right)$ correspond to those of the hole.

Besides these additional diagonal terms, the SW transformation also results in the emergence of off-diagonal terms in $\widetilde{U}$. However, these off-diagonal terms are weak thus can be removed through a second SW transformation $e^{\hat{S}'}\widetilde{U}e^{-\hat{S}'}$, with an anti-hermitian operator $\hat{S}'$ in the following form

$$\hat{S}' = i\begin{pmatrix} 0 & \frac{\alpha}{\epsilon^2_{rc}}\partial_{e,-}U & \frac{\gamma}{\epsilon^2_{rv}}\partial_{e,+}U \\ \frac{\alpha}{\epsilon^2_{rc}}\partial_{e,+}U & 0 & -\frac{\beta}{\epsilon^2_{cv}}\partial_{e,-}U \\ \frac{\gamma}{\epsilon^2_{rv}}\partial_{e,-}U & -\frac{\beta}{\epsilon^2_{cv}}\partial_{e,+}U & 0 \end{pmatrix} \otimes \hat{I}_h$$
$$+\hat{I}_e \otimes i\begin{pmatrix} 0 & -\frac{\alpha'}{\epsilon^2_{r'c'}}\partial_{h,\tau}U & -\frac{\gamma'}{\epsilon^2_{r'v'}}\partial_{h,-\tau}U \\ -\frac{\alpha'}{\epsilon^2_{r'c'}}\partial_{h,-\tau}U & 0 & \frac{\beta'}{\epsilon^2_{c'v'}}\partial_{h,\tau}U \\ -\frac{\gamma'}{\epsilon^2_{r'v'}}\partial_{h,\tau}U & \frac{\beta'}{\epsilon^2_{c'v'}}\partial_{h,-\tau}U & 0 \end{pmatrix}. \quad (A4)$$

The final form of the exciton Hamiltonian $e^{\hat{S}'}e^{\hat{S}}\widehat{H}e^{-\hat{S}}e^{-\hat{S}'}$ becomes fully diagonal up to the first-order of $U$, since $\hat{S}'$ has negligible effects on $\widetilde{H}_{eh}$ due to the large inter-band energy separations. The exciton Hamiltonian in the subspace with the electron located in $c$-band and hole in $v'$-band is

$$\widehat{H}_{cv'} = \epsilon_{cv'} + \frac{\hat{\mathbf{p}}^2}{2m_c} + \frac{\hat{\mathbf{k}}^2}{2m_{v'}} + U(\mathbf{r}_e, \mathbf{r}_h) + \widehat{H}^{(c)}_{\text{Darwin}} + \widehat{H}^{(v')}_{\text{Darwin}} + \widehat{H}^{(c)}_{\text{SOC}} + \widehat{H}^{(v')}_{\text{SOC}}.$$

This then leads to the effective exciton Hamiltonian of Eq. (7).

**Appendix II. Numerically solved exciton wave functions**

We obtained the wave functions $\psi_{nl}(\mathbf{r}) \equiv \langle \mathbf{r}|nl\rangle$ of the exciton Rydberg states by numerically solving the Schrodinger equation

$$\left[\frac{\hbar^2\hat{\mathbf{q}}^2}{2\mu} + V(r)\right]\psi_{nl}(\mathbf{r}) = E_{nl}\psi_{nl}(\mathbf{r}). \quad (A5)$$

We take the values of reduced mass $\mu = 0.31 m_0$ and exciton mass $M = 1.27 m_0$ with $m_0$ the free electron mass, which correspond to the case of homobilayer or monolayer MoSe$_2$. In monolayer TMDs, the e-h Coulomb potential is taken as the Rytova-Keldysh potential [50]

$$V_{\text{monolayer}}(r) = -\frac{e^2}{8\varepsilon\varepsilon_0 r_0}\left[H_0\left(\frac{r}{r_0}\right) - Y_0\left(\frac{r}{r_0}\right)\right], \quad (A6)$$

where $r_0 \approx 3.9/\varepsilon$ nm is the screening length and $\varepsilon$ is the average environmental dielectric constant ($\varepsilon = 1$ for suspended TMDs, and $\varepsilon = 4.5$ for hBN-encapsulated TMDs). While for the interlayer exciton in bilayer TMDs, the e-h Coulomb interaction is [51]

$$V_{\text{interlayer}}(r) = -\frac{e^2}{4\pi\varepsilon\varepsilon_0}\int_0^\infty dq \frac{J_0(rq)}{(1+r_0 q)^2 e^{dq} - r_0^2 q^2 e^{-dq}}. \tag{A7}$$

with $J_0(x)$ as zero order Bessel function of the first kind, and $d \approx 0.6$ nm is the interlayer distance. For $d \to 0$, $V_{\text{interlayer}}$ becomes $V_{\text{monolayer}}$ with a screening length $2r_0$.

The numerically solved wave functions for $1s$, $2s$ and $2p_\pm$ states are shown in Fig. A1, which deviate from the traditional exponentially decaying forms. We find that the numerical results can be fit nearly perfectly by

$$\begin{aligned}\psi_{1s}(\mathbf{r}) &= \frac{1}{\sqrt{2\pi}}\beta_{1s} e^{-\frac{\beta_{1s}r^2}{a+r}}, \\ \psi_{2s}(\mathbf{r}) &= \frac{1}{\sqrt{2\pi}}b_{2s}(1+cr^2)e^{-\frac{\beta_{2s}r^2}{b+r}}, \\ \psi_{2p_\pm}(\mathbf{r}) &= \frac{1}{\sqrt{2\pi}}b_{2p}e^{\pm i\theta}re^{-\frac{\beta_{2p}r^2}{a+r}}.\end{aligned} \tag{A8}$$

The fitting results for excitons in hBN-encapsulated monolayer MoSe$_2$ are shown in Fig. A1.

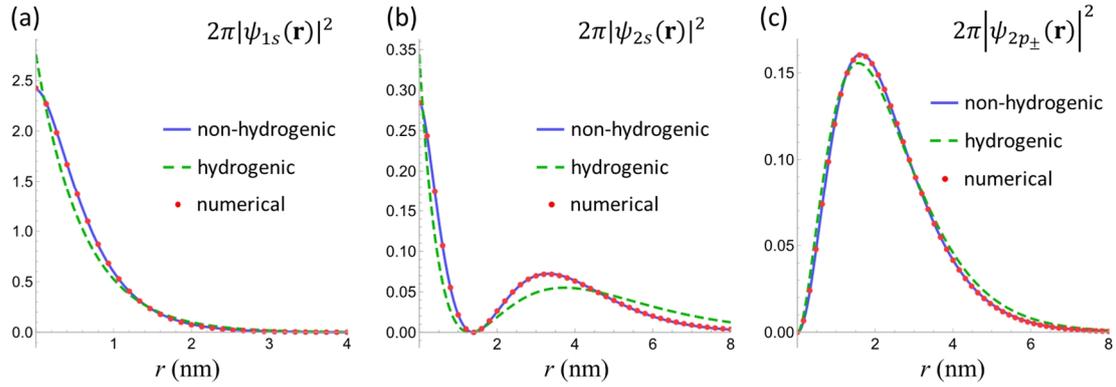

Figure A1. (a), (b), (c) Fitting results for wave functions $|\psi_{1s}|^2$, $|\psi_{2s}|^2$ and $|\psi_{2p_\pm}|^2$, respectively, for excitons in hBN-encapsulated monolayer MoSe$_2$ with $\varepsilon = 4.5$ and $r_0 = 3.9/\varepsilon$ nm. The Blue lines are the fitting results using non-hydrogenic functions $\propto f(r,\theta)e^{-br^2/(a+r)}$ (Eq. (A8)). The green lines are the fitting results using exponentially decaying functions $\propto f(r,\theta)e^{-br}$. The dotted lines are the numerical results.

The energy corrections on the Rydberg states from the Darwin and SOC terms can then be calculated:

$$\begin{aligned}E_{nl}^{(\text{Darwin})} &= A\langle\psi_{nl}|\partial^2 V|\psi_{nl}\rangle, \\ E_{nl}^{(\text{SOC})} &= \frac{\Omega_c + \tau'\Omega_{v'}}{2}\langle\psi_{nl}|(\frac{\partial V}{\partial \mathbf{r}} \times \hat{\mathbf{q}})\cdot \mathbf{e}_z|\psi_{nl}\rangle.\end{aligned} \tag{A9}$$

Note that $V(r)|_{r\to 0} \to \frac{e^2}{4\pi\varepsilon\varepsilon_0 r_0}\ln\left(\frac{r}{2r_0}\right)$, which results in $\partial^2 V(r)|_{r\to 0} \to \frac{e^2}{2\varepsilon\varepsilon_0 r_0}\delta(\mathbf{r})$ and contributes a component $A\frac{e^2}{2\varepsilon\varepsilon_0 r_0}|\psi_{nl}(\mathbf{r}=0)|^2$ to $E_{nl}^{(\text{Darwin})}$. Meanwhile $\partial^2 V(r) -$

$\frac{e^2}{2\epsilon\epsilon_0 r_0}\delta(\mathbf{r}) \neq \mathbf{0}$ when away from $\mathbf{r} = \mathbf{0}$, which contributes a second component to $E_{nl}^{(\text{Darwin})}$. The overall value of $E_{nl}^{(\text{Darwin})}$ is the sum of these two contributions.

1. Ciarrocchi, A., et al., Excitonic devices with van der Waals heterostructures: valleytronics meets twistronics, *Nat. Rev. Mater.* 7(6), 449-464 (2022)
2. Mak, K.F., D. Xiao, and J. Shan, Light-valley interactions in 2D semiconductors, *Nat. Photonics* 12(8), 451-460 (2018)
3. Xu, X.D., et al., Spin and pseudospins in layered transition metal dichalcogenides, *Nat. Phys.* 10(5), 343-350 (2014)
4. Mak, K.F. and J. Shan, Photonics and optoelectronics of 2D semiconductor transition metal dichalcogenides, *Nat. Photonics* 10(4), 216-226 (2016)
5. Wang, Q.H., et al., Electronics and optoelectronics of two-dimensional transition metal dichalcogenides, *Nat. Nanotechnol.* 7(11), 699-712 (2012)
6. Chernikov, A., et al., Exciton Binding Energy and Nonhydrogenic Rydberg Series in Monolayer WS2, *Phys. Rev. Lett.* 113(7), 076802 (2014)
7. Qiu, D.Y., F.H. da Jornada, and S.G. Louie, Optical Spectrum of MoS2: Many-Body Effects and Diversity of Exciton States, *Phys. Rev. Lett.* 111(21), 216805 (2013)
8. Yang, X.L., et al., Analytic solution of a two-dimensional hydrogen atom. I. Nonrelativistic theory, *Phys. Rev. A* 43(3), 1186-1196 (1991)
9. Cao, T., et al., Valley-selective circular dichroism of monolayer molybdenum disulphide, *Nat. Commun.* 3(1), 887 (2012)
10. Zeng, H.L., et al., Valley polarization in MoS2 monolayers by optical pumping, *Nat. Nanotechnol.* 7(8), 490-493 (2012)
11. Mak, K.F., et al., Control of valley polarization in monolayer MoS2 by optical helicity, *Nat. Nanotechnol.* 7(8), 494-498 (2012)
12. Jones, A.M., et al., Optical generation of excitonic valley coherence in monolayer WSe, *Nat. Nanotechnol.* 8(9), 634-638 (2013)
13. Rivera, P., et al., Interlayer valley excitons in heterobilayers of transition metal dichalcogenides, *Nat. Nanotechnol.* 13(11), 1004-1015 (2018)
14. Xiao, D., M.C. Chang, and Q. Niu, Berry phase effects on electronic properties, *Rev. Mod. Phys.* 82(3), 1959-2007 (2010)
15. Xiao, D., et al., Coupled Spin and Valley Physics in Monolayers of MoS2 and Other Group-VI Dichalcogenides, *Phys. Rev. Lett.* 108(19), 196802 (2012)
16. Aivazian, G., et al., Magnetic control of valley pseudospin in monolayer WSe2, *Nat. Phys.* 11(2), 148-152 (2015)
17. Srivastava, A., et al., Valley Zeeman effect in elementary optical excitations of monolayer WSe, *Nat. Phys.* 11(2), 141-147 (2015)
18. MacNeill, D., et al., Breaking of Valley Degeneracy by Magnetic Field in Monolayer MoSe2, *Phys. Rev. Lett.* 114(3), 037401 (2015)
19. Kormanyos, A., et al., Tunable Berry curvature and valley and spin Hall effect in bilayer MoS2,

*Phys. Rev. B* 98(3), 035408 (2018)

20. Mak, K.F., et al., The valley Hall effect in MoS2 transistors, *Science* 344(6191), 1489-1492 (2014)
21. Yu, T. and M.W. Wu, Valley depolarization dynamics and valley Hall effect of excitons in monolayer and bilayer MoS2, *Phys. Rev. B* 93(4), 045414 (2016)
22. Yao, W., D. Xiao, and Q. Niu, Valley-dependent optoelectronics from inversion symmetry breaking, *Phys. Rev. B* 77(23), 235406 (2008)
23. Zhu, Q.Z., et al., Gate tuning from exciton superfluid to quantum anomalous Hall in van der Waals heterobilayer, *Sci. Adv.* 5(1), eaau6120 (2019)
24. Jiang, C.Y., et al., A room-temperature gate-tunable bipolar valley Hall effect in molybdenum disulfide/tungsten diselenide heterostructures, *Nat. Electron.* 5(1), 23-27 (2022)
25. Onga, M., et al., Exciton Hall effect in monolayer MoS2, *Nat. Mater.* 16(12), 1193-1197 (2017)
26. Huang, Z., et al., Robust room temperature valley Hall effect of interlayer excitons, *Nano Lett.* 20(2), 1345-1351 (2019)
27. Yao, W. and Q. Niu, Berry phase effect on the exciton transport and on the exciton Bose-Einstein condensate, *Phys. Rev. Lett.* 101(10), 106401 (2008)
28. Yu, H.Y. and W. Yao, Electrically tunable topological transport of moire polaritons, *Sci. Bull.* 65(18), 1555-1562 (2020)
29. Ubrig, N., et al., Microscopic Origin of the Valley Hall Effect in Transition Metal Dichalcogenides Revealed by Wavelength-Dependent Mapping, *Nano Lett.* 17(9), 5719-5725 (2017)
30. Trushin, M., M.O. Goerbig, and W. Belzig, Model Prediction of Self-Rotating Excitons in Two-Dimensional Transition-Metal Dichalcogenides, *Phys. Rev. Lett.* 120(18), 187401 (2018)
31. Hichri, A., S. Jaziri, and M.O. Goerbig, Charged excitons in two-dimensional transition metal dichalcogenides: Semiclassical calculation of Berry curvature effects, *Phys. Rev. B* 100(11), 115426 (2019)
32. Srivastava, A. and A. Imamoğlu, Signatures of bloch-band geometry on excitons: Nonhydrogenic spectra in transition-metal dichalcogenides, *Phys. Rev. Lett.* 115(16), 166802 (2015)
33. Zhou, J.H., et al., Berry Phase Modification to the Energy Spectrum of Excitons, *Phys. Rev. Lett.* 115(16), 166803 (2015)
34. Gong, P., et al., Optical selection rules for excitonic Rydberg series in the massive Dirac cones of hexagonal two-dimensional materials, *Phys. Rev. B* 95(12), 125420 (2017)
35. Cao, T., M. Wu, and S.G. Louie, Unifying optical selection rules for excitons in two dimensions: band topology and winding numbers, *Phys. Rev. Lett.* 120(8), 087402 (2018)
36. Zhang, X.O., W.Y. Shan, and D. Xiao, Optical Selection Rule of Excitons in Gapped Chiral Fermion Systems, *Phys. Rev. Lett.* 120(7), 077401 (2018)
37. Liu, G.B., et al., Three-band tight-binding model for monolayers of group-VIB transition metal dichalcogenides, *Phys. Rev. B* 88(8), 085433 (2013)
38. Wu, F.C., F.Y. Qu, and A.H. MacDonald, Exciton band structure of monolayer MoS2, *Phys. Rev. B* 91(7), 075310 (2015)
39. Ye, Z.L., et al., Probing excitonic dark states in single-layer tungsten disulphide, *Nature* 513(7517), 214-218 (2014)
40. Qiu, D.Y., T. Cao, and S.G. Louie, Nonanalyticity, Valley Quantum Phases, and Lightlike

Exciton Dispersion in Monolayer Transition Metal Dichalcogenides: Theory and First-Principles Calculations, *Phys. Rev. Lett.* 115(17), 176801 (2015)

41. Yong, C.K., et al., Valley-dependent exciton fine structure and Autler-Townes doublets from Berry phases in monolayer MoSe2, *Nat. Mater.* 18(10), 1065-1070 (2019)
42. Chaudhary, S., C. Knapp, and G. Refael, Anomalous exciton transport in response to a uniform in-plane electric field, *Phys. Rev. B* 103(16), 165119 (2021)
43. Cao, J.L., H.A. Fertig, and L. Brey, Quantum geometric exciton drift velocity, *Phys. Rev. B* 103(11), 115422 (2021)
44. Sui, M.Q., et al., Gate-tunable topological valley transport in bilayer graphene, *Nat. Phys.* 11(12), 1027-1031 (2015)
45. Shimazaki, Y., et al., Generation and detection of pure valley current by electrically induced Berry curvature in bilayer graphene, *Nat. Phys.* 11(12), 1032-1036 (2015)
46. Ju, L., et al., Tunable excitons in bilayer graphene, *Science* 358(6365), 907-910 (2017)
47. Yu, H.Y. and W. Yao, Luminescence Anomaly of Dipolar Valley Excitons in Homobilayer Semiconductor Moire Superlattices, *Phys. Rev. X* 11(2), 021042 (2021)
48. Yu, H.Y., et al., Dirac cones and Dirac saddle points of bright excitons in monolayer transition metal dichalcogenides, *Nat. Commun.* 5(1), 3876 (2014)
49. He, M.H., et al., Valley phonons and exciton complexes in a monolayer semiconductor, *Nat. Commun.* 11(1), 618 (2020)
50. Cudazzo, P., I.V. Tokatly, and A. Rubio, Dielectric screening in two-dimensional insulators: Implications for excitonic and impurity states in graphane, *Phys. Rev. B* 84(8), 085406 (2011)
51. Danovich, M., et al., Localized interlayer complexes in heterobilayer transition metal dichalcogenides, *Phys. Rev. B* 97(19), 195452 (2018)